\documentclass{JHEP3}
\usepackage{graphicx}
\usepackage{epsfig}
\usepackage{amsmath}
\usepackage{bbm}
\usepackage{cite}

\newcommand{\bee}{\begin{equation}}
\newcommand{\ee}{\end{equation}}
\newcommand{\beea}{\begin{eqnarray}}
\newcommand{\eea}{\end{eqnarray}}
\newcommand{\gfive}{\gamma_5}

\def\Tr{{\rm Tr}}

\title{Simulating an arbitrary number of flavors of dynamical overlap fermions}

\author{Thomas DeGrand$^a$, Stefan Schaefer$^b$ \\
\it $^a$Department of Physics, University of Colorado, Boulder, CO 80309 USA \\
\it $^b$NIC, DESY Zeuthen, Platanenallee 6, D-15738 Zeuthen, Germany
}

\date{\today}
\abstract{
We present a set of related
 Hybrid Monte Carlo methods to simulate an arbitrary number of dynamical
 overlap fermions.
Each fermion is represented by a chiral pseudo-fermion field. The new algorithm
reduces critical slowing down in the chiral limit and for sectors of nontrivial
topology.
}

\preprint{DESY 06-056}

\begin{document}

\section{Introduction}
Contemporary simulations of full QCD on the lattice typically use two degenerate light 
flavors ('up' and 'down') and one heavier flavor ('strange') of sea quarks.
For each flavor the fermionic determinant is replaced by an integral over  bosonic fields $\phi$,
the pseudofermions \cite{Weingarten:1980hx}
\bee
\det D \propto \int [d \phi^\dagger ][d \phi] e^{-\phi^\dagger  \frac{1}{D} \phi} \ .
\label{eq:pf1}
\ee
This identity  requires all eigenvalues of the matrix $D$ to have a  positive real part.

Most lattice Dirac operators obey $\gamma_5$-Hermiticity, $D^\dagger  = \gamma_5 D \gamma_5$,
and so $\det D$ is real. However, the eigenvalues of lattice Dirac operators
are typically complex and their real
parts may not be positive-definite.  Then the exponential in
Eq.~(\ref{eq:pf1}) cannot be interpreted as a conventional probability
measure.
This is the case for Wilson fermion actions. They have  eigenvalues that are paired
complex-conjugates or unpaired and real:
\bee
\det D =
\prod_{\rm pairs}(|\lambda_j|^2 + m^2) \prod_r(\lambda_r + m) \ .
\label{eq:pfx}
\ee
The absence of chiral symmetry
means that the sign of the real eigenvalues is not protected, so the
determinant can have either sign. 
The solution to this problem is to simulate two degenerate flavors at a time, that is, to rewrite
their fermion determinant
$\det^2 D = \det D^\dagger D$ where $D^\dagger D$ fulfills the requirements of Eq.~\ref{eq:pf1}.
So Eq.~\ref{eq:pfx}  becomes
\bee
\det D^\dagger  D = \prod_i(|\lambda_i|^2 + m^2) \prod_r (\lambda_r + m)^2 \ .
\ee
where the product index $i$ runs over all modes. Simulations of  a single flavor are, in general, 
not as straightforward, and one is forced to use either the Refreshed Molecular Dynamics or R
algorithm\cite{Gottlieb:1987mq}, which is not exact, or the
Rational Hybrid Monte Carlo (RHMC) algorithm\cite{Clark:2004cp}, which uses the
 square root of the two flavor 
operator, $Q=\sqrt{D^\dagger D}$. This could introduce a systematic error if
$\lambda_r+m$ could change sign.

 For the overlap Dirac operator 
\cite{Neuberger:1997fp,Neuberger:1998my}
there is a more natural solution.
It exploits the chiral properties of the formulation, i.e. that the eigenmodes
of $D^\dagger D$ can be chosen to be chiral,
 with an eigenmode of each chirality per eigenvalue $|\lambda_i|^2+m^2$.
A single flavor of overlap fermions can be simulated with one chiral pseudofermion.
 One only has to correct for
the effect of the real modes. The real spectrum is known if one knows the topological 
charge of the configuration as defined by the index of the Dirac operator. 
The idea is based on a suggestion
which can already be found in Ref.~\cite{Bode:1999dd} but shall now be formulated 
and investigated more precisely.

A simulation of any number $n_f$ of flavors  thus
involves a set of $n_f$ chiral pseudo-fermion fields.
It is also necessary to keep track of the global topology of the configuration during the simulation.
Besides allowing simulations for any $n_f$, the chiral algorithm has an advantage over the
traditional ``non-chiral''  Hybrid Monte Carlo (HMC)
 algorithm (with a Dirac spinor pseudo-fermion for each degenerate flavor pair):
Away from topological boundaries, or  in simulations in which topological changes are
forbidden, it can eliminate  part of the critical 
slowing down in the chiral limit and  stabilize the inversion in  sectors
of non-trivial topology. This is achieved simply by running in the sector of chirality which
is topologically trivial.

In Sec. 2 we describe the method. The new ingredient, not mentioned in Ref. \cite{Bode:1999dd},
is a technique for generating the initial chiral pseudofermion. In Sec. 3 we present
results of some test simulations in small volumes. Some of these results are surprising, so in
Sec. 4 we describe a little solvable model which reproduces these features.

\section{The method}
Let us first fix our conventions and repeat a few properties of Neuberger's overlap
operator. It is given by
\bee
D =D_{ov}(m=0)= R_0 \left[ 1+\gfive \epsilon(h(-R_0))\right]
\label{eq:Dov}
\ee
with $\epsilon(h)=h/\sqrt{h^2}$ the sign function of the Hermitian kernel
operator $h=\gfive d$ which is taken at the negative mass $R_0$.
Its spectrum  is symmetric, i.e. each non-real eigenvalue 
$\lambda$ is paired with its complex conjugate $\lambda^*$. 
The modes at zero and $2R_0$ are the only real modes. They are chiral and the 
excess of the zero modes of negative chirality over the ones with positive chirality
gives the topological charge Q. Experience shows that there
are always only zero modes of one chirality. For  simplicity of the argument
we will assume this in the following 
even though this assumption is not necessary.
The squared Hermitian overlap operator 
$H^2=(\gamma_5 D)^2=D^\dagger D $
commutes with $\gamma_5$ and therefore can have eigenvectors with 
definite chirality. The modes at zero and $4R_0^2$ aside, the spectrum is doubled 
with a positive and a negative chirality eigenvector to each eigenvalue
$|\lambda|^2$.
It is therefore convenient to define the chiral projections ($P_\pm = \frac{1}{2}(1\pm \gamma_5)$)
so that the massive squared Hermitian overlap operator, with the usual
convention for the mass terms, is
\bee
H^2_\pm(m) =  P_\pm H^2(m) P_\pm=2(R_0^2-\frac{m^2}{4}) P_\pm (1+\epsilon(h))P_\pm+m^2 P_\pm \ .
\ee

Let us call the chiral sector without the zero modes the sector 
of opposite chirality and the associated Hermitian Dirac operator 
$H^2_{\rm opp}$.
The chiral sector with the chirality of the zero modes shall be 
called the same chirality sector and $H^2_{\rm same}$ the associated operator.

The method which we are going to discuss  exploits the fact that $H^2$ has the same 
spectrum in both chiral sectors. Only the modes at $m^2$ and $4R_0^2$ differ. The 
fermion determinant for one flavor is therefore the determinant of $H^2$ 
in one chiral sector times a correction factor for the modes at $m$ and $2R_0$. 
This is summarized in the following relation
\bee
{\det} D = (m/2R_0)^{|Q|} {\det} H_{\rm opp}^2
= (2R_0/m)^{|Q|} {\det} H_{\rm same}^2 \ . 
\ee
Since  $H_{\rm opp}^2$ and  $H_{\rm same}^2$ are positive operators, it is thus 
straightforward to simulate a single flavor in hybrid Monte Carlo.

First, however, one has to decide  which action  to simulate, i.e. how to choose the
optimal chirality for the pseudofermion
 a given gauge configuration. To begin, for any 
configuration with topological charge, the inversion of the Dirac operator
is computationally cheaper in the sector of opposite chirality.
Typically, the non-zero modes are repelled by the 
zero mode(s) and the smallest non-zero mode is significantly above zero. The 
conditioning number of the Dirac operator (the ratio of largest to smallest 
eigenvalue) is therefore  lower in the opposite chirality sector
because the largest eigenvalue is always
close to $2R_0$ regardless of chirality. (For $H(m)^2$ it is proportional to
$1/(m^2 + \lambda_{min}^2)$.)

During the molecular dynamics evolution the topological charge can change.
One has to decide what to do when the chirality of the zero-modes flips (e.g. starting in
a trajectory
 in a configuration with $Q=1$ and evolving through a region with $Q=0$ into $Q=-1$).
There are in principle two options: One can keep the chirality of
the operator to be fixed, or one can choose the chirality
of the operator to be in the opposite chirality sector of the configuration. 
This means that when one changes the topology of the configuration, one also
changes the chirality of the pseudofermion.
The associated action is given by
\bee
\det D=(m/2R_0)^{|Q|}{\det} H_{\rm opp}^2 \propto
\int [d\phi][d\phi^\dagger]  \exp(|Q| \log \frac{m}{2R_0} - \phi^\dagger  H_{\rm opp}^{-2} \phi) \ . \label{eq:act1}
\ee

To simulate this action we use the Hybrid Monte Carlo algorithm. We build on experiences
previously published in Refs.~\cite{DeGrand:2004nq,DeGrand:2005vb,Schaefer:2005qg}. The
initial formulation of the HMC algorithm \cite{Duane:1987de} 
for the overlap operator is given in Ref.~\cite{Fodor:2003bh}.
At the start of a trajectory, one has to perform a heat-bath for the pseudo-fermion
fields. In a conventional $n_f=2$ simulation, one would cast a random vector $\xi$ in both chiralities
and compute $\phi = H \xi$. Then one would break $\phi$ into its separate chiralities
and use
\bee
\det D^2 = \int [d\phi_+][d\phi_+^\dagger]  [d\phi_-][d\phi_-^\dagger ]
 \exp(-\phi_+^\dagger  H_+^{-2} \phi_+ -\phi_-^\dagger  H_-^{-2} \phi_-).
\ee
However, we need a set of chiral $\phi$ fields, of chirality $\sigma$, chosen by heat bath.
(How to do this is not described explicitly  in Ref. \cite{Bode:1999dd}.)
 We achieve this goal
by generating a set of chiral Gaussian random fields $\xi_\sigma$. We then define
$\phi_\sigma = \sqrt{H_\sigma^2} \xi_\sigma$.
To construct $\phi_\sigma$ we use a rational approximation to the square root in the region
$[m^2,4 R_0^2]$, i.e. we approximate it by  
\bee
\sqrt{H_\sigma^2} \xi_\sigma \approx H_\sigma^2 \sum_l \frac{b_l}{H_\sigma^2 +
c_l} \xi_\sigma
\label{eq:zolstart}
\ee
where we use the  $b_l$ and $c_l$ from the Zolotarov approximation to the sign function (there used
to compute $\epsilon(x)=x/\sqrt{x^2}$). 

When we use Hasenbusch preconditioning\cite{Hasenbusch:2001ne}, the action for the
lower mass pseudofermions is
\bee
S_f = \phi_\sigma^\dagger \frac{H_\sigma(m')^2}{H_\sigma(m)^2} \phi_\sigma .
\ee
Our chiral pseudofermion is then taken to be $\sqrt{H_\sigma(m)^2/H_\sigma(m')^2}
\xi_\sigma$,
which we again approximate by a Zolotarov formula, whose independent variable $x$ obeys
\bee
x^{-1} = H_\sigma(m')^2/H_\sigma(m)^2 = \alpha + \beta/H_\sigma(m)^2 .
\ee
The range of $x$ is $(m/m')^2$ to 1.

During the trajectory one has to deal with the discontinuity due to the sign function
in the definition of the overlap operator, when an eigenmode  $|\lambda_0\rangle$
of the kernel operator $h(-R_0)$ changes
sign.
Fodor et al. \cite{Fodor:2003bh} proposed a method of how to deal with this problem.
 One  measures
the height of the step in the effective action and then reflects or refracts the gauge field momentum
as in classical mechanics. 
The computation of the height of the step is a major part of the total cost of the
simulation.  If we use the same chirality of $H_\sigma$ on both sides of that boundary this
amounts to the change
\bee
H^2_\sigma(m)
 \longrightarrow H^2_\sigma(m)\pm(4R_0^2-m^2) P_\sigma|\lambda_0\rangle\langle \lambda_0|P_\sigma 
\equiv  \tilde H^2_\sigma(m)\ .
\label{eq:dspf}
\ee
From the Sherman-Morrison formula,
\bee
\frac{1}{\tilde H^2_\sigma(m)} = \frac{1}{ H^2_\sigma(m)} - \frac{\delta C}{1 +\delta C L}
\frac{1}{ H^2_\sigma(m)} P_\sigma|\lambda_0 \rangle\langle \lambda_0|P_\sigma
\frac{1}{ H^2_\sigma(m)} ,
\label{eq:sm}
\ee
so the height of the step is given by
\bee
\Delta \left [ \langle \phi| P_\sigma \frac{1}{\tilde H_\sigma(m)^2}P_\sigma| \phi \rangle \right] =-\frac{\delta C}{1 +\delta C L}
|\langle  \phi |P_\sigma\frac{1}{H_\sigma(m)^2}P_\sigma| \lambda_0\rangle|^2 \ .
\label{eq:sherman}
\ee
Finally, for completeness, the
 exact ratio of determinants is
\bee
  \frac{\det \tilde H^2_\sigma(m)}{\det  H^2_\sigma(m)}
 =   1 + \delta C L . \label{eq:stepdet}
\ee
In the abbreviated formulas $\delta$ is the sign in Eq.~\ref{eq:dspf}, $C = (4R_0^2-m^2)$
 and $L$ is the matrix element 
$\langle \lambda_0|P_\sigma H^{-2}_\sigma(m) P_\sigma|\lambda_0 \rangle$.

Eq.~\ref{eq:sherman} is obviously only applicable if we use the same chiral sector on
both sides. Otherwise, one has to run the inversion twice, which is very expensive and
numerically less under control. 
This will occur during crossings into $Q=0$.
Technically, a change in the chirality which we are using for $H^2_\sigma$ amounts to a
change in the chirality which we use for the embedding of the two-component chiral source
into the four-component Wilson vector. We will therefore use the term source chirality for 
chirality of $H^2$.

Since choosing the chirality which we use in the $Q=0$ sector depending on
the configuration we start the trajectory from would violate
reversibility, we choose it randomly at the beginning of each trajectory.
%
%
So at a topological boundary, where we propose a tunneling into $Q=0$, the new source
chirality will be different from the initial chirality half the time.
 The $\Delta S$ which results from flipping the source chirality
is very large and most of these crossings will be rejected.
This suggests two other algorithms for simulating any number of
flavors:
\begin{itemize}
\item{Simply restrict the simulation to a particular topological sector. There are
simulational situations where this restriction is desirable.
They include the calculation of the condensate using random matrix theory, or
calculations of full QCD in the so-called epsilon regime.
It is unknown whether these simulations are ergodic. If the manifold of gauge fields corresponding to
fixed topology were smoothly connected, then HMC would (in principle) carry us
 from any gauge configuration
to any other one by a series of small steps.
However, if sectors of fixed topology were
 disjoint, or could be connected only by passage a sector of some other topological charge,
HMC in a sector of fixed topology would not be ergodic. We are aware of no proofs one way or the
 other, for four dimensions. Arguments we construct based on instanton phenomenology,
where $Q$ counts the excess of instantons over anti-instantons,
argue that there is no problem: in the different configurations, the location of the odd
 instanton(s) moves around, and their
sizes shrink and grow, but this is all continuous. So is the appearance of pairs
of
instantons and anti-instantons, as they grow from fluctuations of a single plaquette, or
annihilate similarly.  To produce the entire functional integral, results
from different topological sectors can be combined using Eq. \ref{eq:stepdet},
as described by Ref. \cite{Egri:2005cx}.
}
\item{Alter the tunneling probability and reweight the resulting data set, if necessary. A simple
way to do this is to pick all chiral sources to carry the same chirality 
and begin the simulation
either in $Q=0$ or in a topological sector in which the sources do not have zero modes.
Here we must assume that configurations with zero modes in both chiralities never
appear.
Then allow topological changes  in which the sources
do not have zero modes, but prohibit transitions which would create zero
 modes in the source chirality.
For example, we could set the source chiralities to be positive and only allow transitions
into $Q=n_- -n_+\ge 0$. (We will call this the ``fixed chirality algorithm.'')
 Unless there are disconnected sectors at $Q \ne 0$ which can only
be reached by some passage through 
$Q<0$ (for example $Q=1 \rightarrow Q=0 \rightarrow Q=-1 \rightarrow Q=0 \rightarrow Q=1
\rightarrow Q=2$) the algorithm
 will generate an ensemble with the correct Boltzmann weighting ratio between sectors
of all $Q\ge 0$. Under a parity transformation a gauge configuration with positive 
$Q$ is converted
into one with negative $Q$. In the analysis of an ensemble generated with
this algorithm, measurements
on the $Q=0$ configurations need to be reweighted with a factor $1/2$
compared to
those from configurations with non-trivial topology.

At a topological boundary we add  $\Delta|Q|\log(m/(2R_0))$ to the pseudofermion
$\Delta S$ before deciding whether to reflect or refract.  As a variant on this
approach (which we have not tried) one could leave out a fraction
 of the  $|Q|\log(m/(2R_0))$ factor
from the action during the HMC evolution and include it later with a real reweighting.
}
\end{itemize}

\section{QCD simulations}

We make tests on four different data sets: Set ${\bf A}$ is generated with 
the  standard two-flavor HMC algorithm 
on an $8^3 \times 6$ lattice, at a lattice spacing of  $a\approx0.16$ fm and 
a bare quark mass of $am=0.05$. This is close to the crossover between the
chirally broken low temperature phase and the chirally restored high temperature phase.
The details of this simulation and its parameters are similar to those
discussed in Ref.~\cite{DeGrand:2005vb}. 
We only mention that our gauge connections are stout links \cite{Morningstar:2003gk}  and that
we use Hasenbusch preconditioning\cite{Hasenbusch:2001ne} with one extra pair of 
pseudo-fermion fields at a higher mass. 
Here we  use three levels of stout smearing. 
The molecular dynamics integration has a multiple time step integration similar
 to that of Ref.~\cite{Urbach:2005ji}.
Set ${\bf A'}$ is run at the same parameters but using the new chiral algorithm.
Set ${\bf B}$ is generated with the fixed chirality algorithm.
We restrict these studies to $n_f=2$ because
we can make comparisons to the usual (nonchiral) HMC algorithm. (We have also done extensive
running with $n_f=1$, which we will report elsewhere.)

Finally, we have run the new algorithm to generate sets (labeled ${\bf C}$) 
of $10^4$ lattices at the same lattice spacing as sets ${\bf A}$ and ${\bf A'}$,
with two steps of stout smearing, at fermion 
masses of $am_q=0.05$, 0.03 and 0.015.
 These simulations are done in sectors of fixed topology by switching off the possibility
 of refraction. This allows us to study the behavior in the different topological sectors.
Otherwise, in particular at small quark masses, the fermion determinant suppresses
 the sectors of non-zero topology and it is hard to get  sufficient statistics there.

We check that our starting pseudofermion field is chosen appropriately by
computing $\phi_\sigma^\dagger H_\sigma(m)^{-2} \phi_\sigma$ (or 
when we use Hasenbusch preconditioning,
$\phi_\sigma^\dagger H^2_\sigma(m') H_\sigma(m)^{-2}  \phi_\sigma$) and comparing this value to
the heat bath initialization $\xi^\dagger_\sigma \xi_\sigma$. With high accuracy evaluations of the Zolotarov
formula and a tight convergence criterion  for the Conjugate Gradient inversion of the quark
propagator ($r^\dagger r = 10^{-16})$, the deviation in the action from its heat bath value
is held below 0.02 or so, out of a total fermion energy in our simulations
of a few times $10^5$. This is small compared to the typical violation of energy conservation
in our molecular dynamics trajectory.

Let us turn to the critical slowing down and the cost of the inversion  (we
do not make statements about the auto-correlation time because
our data set is too small to make a definite statement). The new method has
an overhead at the start of the trajectory because one has to hit the Gaussian source
with the square root, which involves a multi-mass inversion of the overlap operator,
instead of just the operator $H$. However, this is only a small fraction of the total
cost of the algorithm.

Since the simulation for set ${\bf C}$  is done in a fixed topological sector, 
we  have significant statistics for the $Q=\pm 1$ sector  for smaller quark masses.
We can thus study the critical slowing down of the inversion at trivial and non-trivial
topology. 
For all three quark masses we found no significant difference in the cost of the inversion 
between the $Q=0$ and the $Q=\pm 1$ runs. At fixed topology, the variation of the 
number of CG steps for the three different bare quark masses ($am=0.015$, $0.03$ and $0.05$)
is below $20\%$.
We can conclude  (at least for the small volumes and for our parameters) that critical slowing
down is largely eliminated by the chiral algorithm. 

The other algorithms allow for topological changes. We made simulation runs of 250-300 trajectories
for each algorithm. We observed that the plaquettes from all simulations were 
consistent within uncertainties. In all three runs we had tunnels from $Q=0$ into and out of
$|Q|=1$. The tunneling rate was too low in all three simulations to say
anything meaningful about autocorrelation times. All three runs used the same parameters.
All had acceptance rates of about 90 per cent. All had about 1.8 attempted 
topological changes per trajectory. The cost of a trajectory in units of the number of applications
of $H_\sigma^2$ to a trial vector were 2430(23) for set ${\bf A}$, 2798(47) for ${\bf A'}$,
and 2221(17) for ${\bf B}$. The excess of ${\bf A'}$ over ${\bf A}$ is
due to the startup. The decrease of ${\bf B}$ from ${\bf A}$ 
is from the modest decrease in the conditioning number
because we never run in a topologically nontrivial sector.

The algorithm refracts when $\Delta S$, the change in action, is smaller than
$\frac{1}{2}\langle N|\pi \rangle^2$, the squared projection of the gauge momentum normal
to the surface of topology change. We first show a histogram of $\langle N|\pi\rangle^2$
in Fig.~\ref{fig:ndoth}. As expected there is little difference in this quantity
 between the  algorithms.

\FIGURE{
\includegraphics[width=0.75\textwidth,clip]{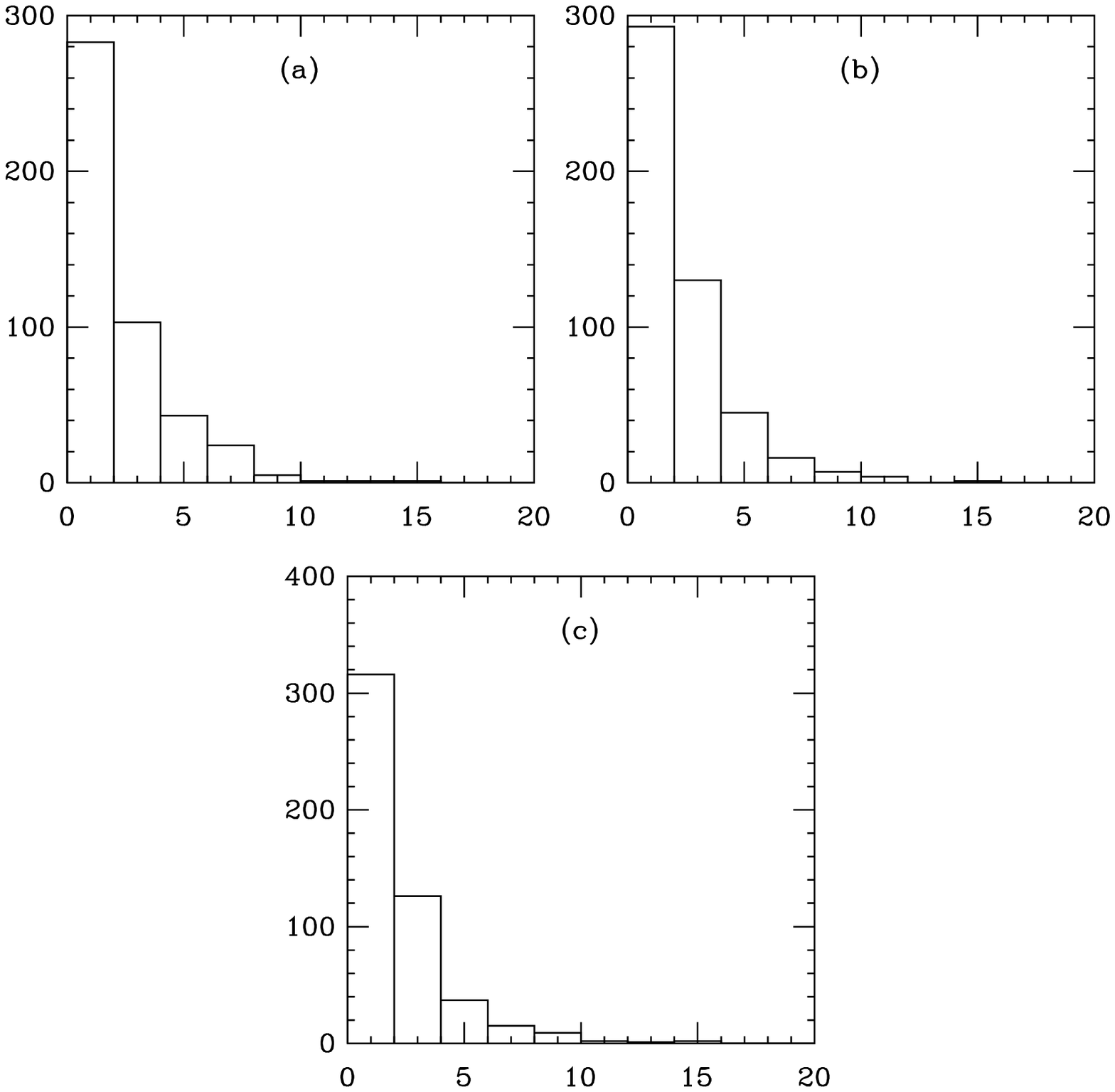}
\caption{Histograms of $\langle N |\pi\rangle^2$, the squared normal component of
the gauge momentum at a topological boundary,
from the non-chiral algorithm ${\bf A}$ (a),  the chiral algorithm ${\bf A'}$ (b), and the
fixed chirality algorithm ${\bf B}$ (c). 
\label{fig:ndoth}
}
}

\FIGURE{
\includegraphics[width=0.75\textwidth,clip]{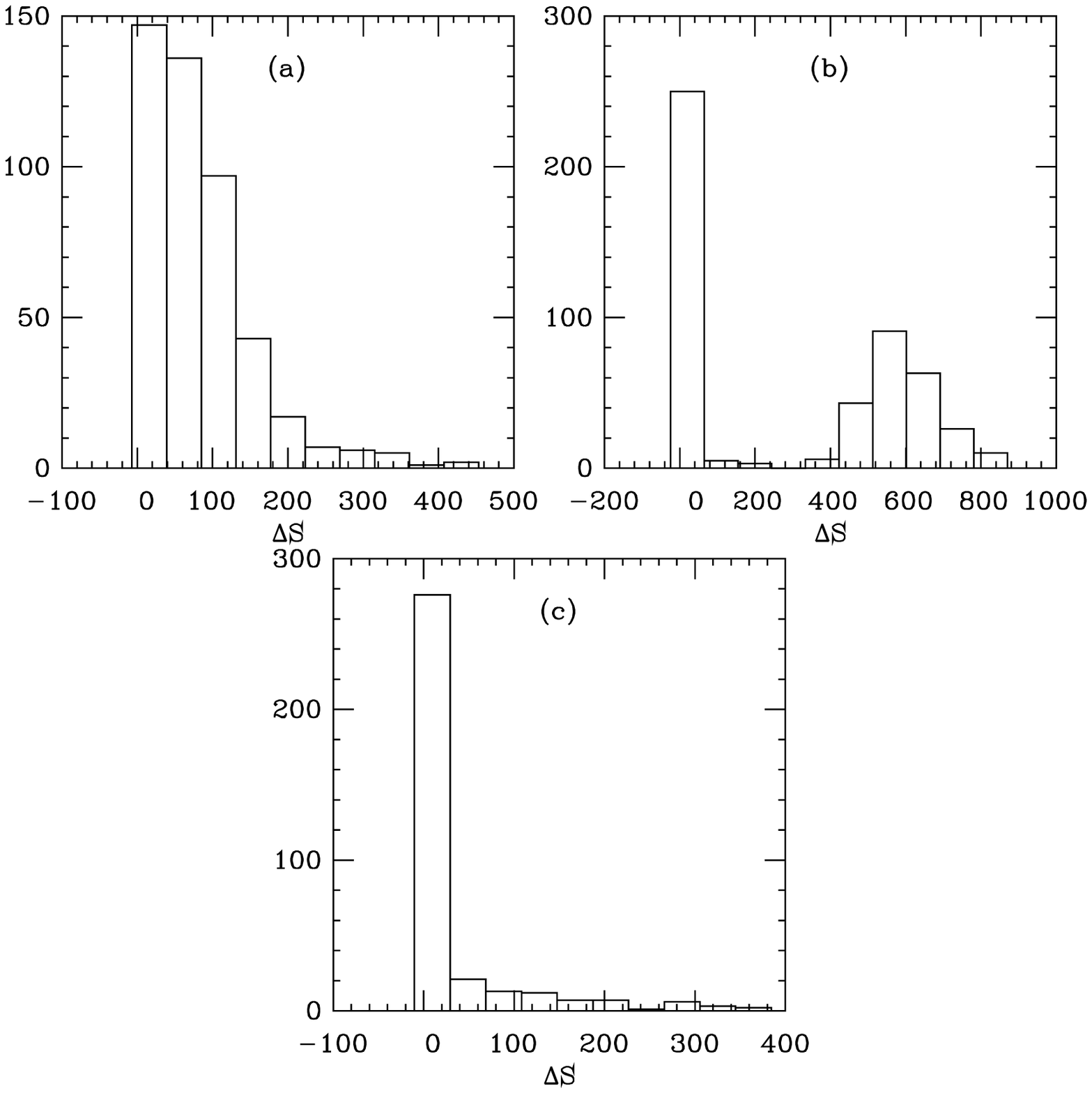}
\caption{Histograms of $\Delta S$, the the height of the step at a topological boundary,
from the non-chiral data set ${\bf A}$ (a),  the chiral data set ${\bf A'}$ (b), and the fixed chirality
data set ${\bf B}$ (c). 
\label{fig:deltas}
}
}
Next we look at $\Delta S$ and show histograms of this quantity in  
Fig.~\ref{fig:deltas}.  These distributions are quite different. To make sense of them,
we realize that while for the nonchiral algorithm all crossings are ``similar,'' in the sense
that all transitions involve contributions from zero modes in either the initial or final
state, that is not the case for the chiral algorithms: either the transition involves
a change in topology in which the zero mode has appeared or disappeared from the opposite
chirality sector, or in the same sector as the simulation. In the case of
an opposite chirality change, the magnitude of the topology can either increase or decrease.

\FIGURE{
\includegraphics[width=0.75\textwidth,clip]{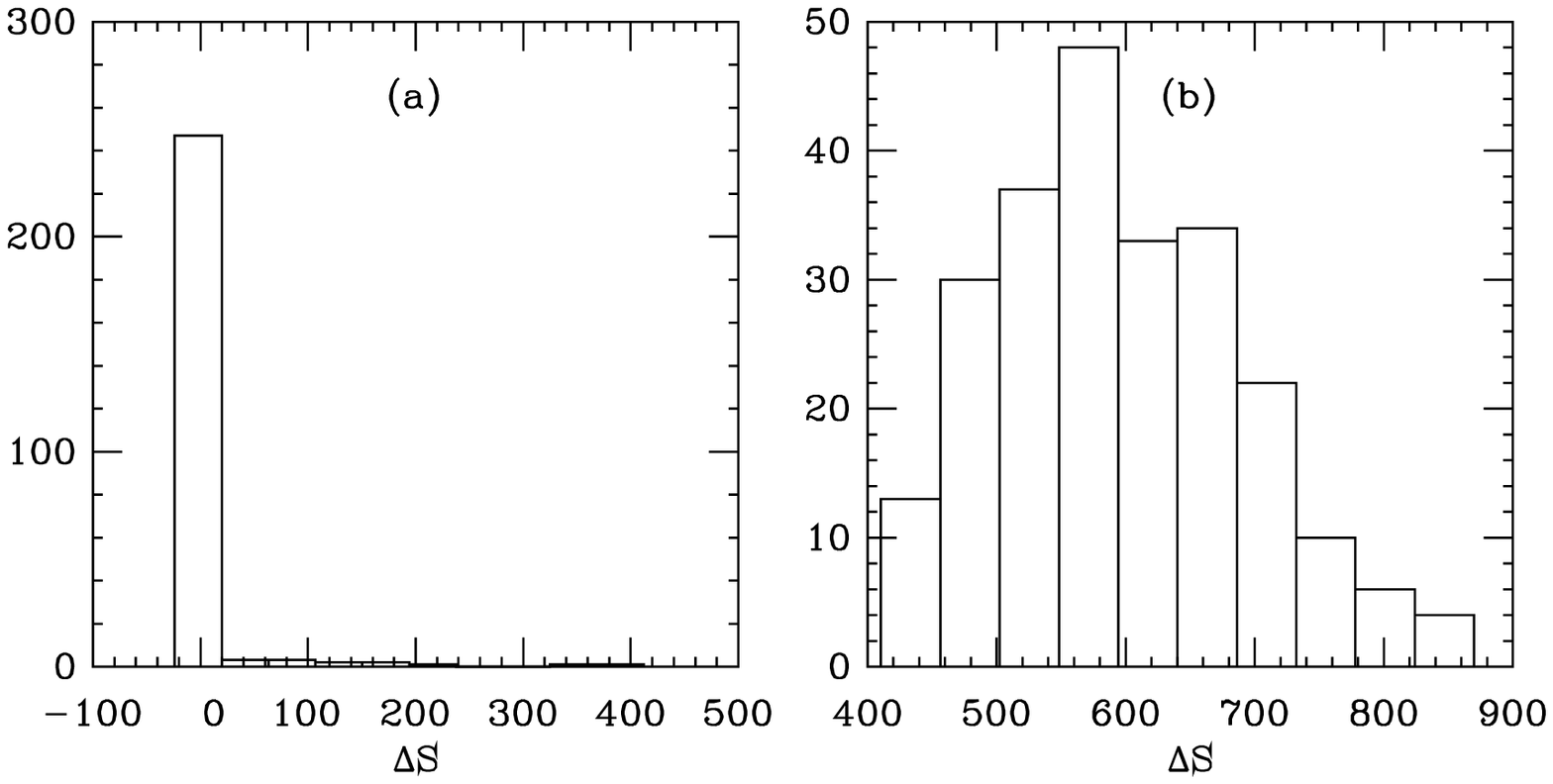}
\caption{Histograms of $\Delta S$ from algorithm ${\bf A'}$, where the proposed topology change is 
from (a) transitions away from the running chirality and (b) into the running chirality.
\label{fig:samediff}
}
}
In Fig. \ref{fig:samediff} we further divide the contributions to  Fig. \ref{fig:deltas}b
and
show histograms of $\Delta S$ from algorithm ${\bf A'}$
for two cases. In panel (a) we show  $\Delta S$ from transitions when the running chirality is
different from the topology of the proposed crossing. This histogram itself
has two components, a narrow one and a wide one, which we will shortly separate.

In panel (b) we show $\Delta S$ for transitions in which the final topology
and the running topology have the same sign. In this algorithm we compute
the action in the new sector by flipping the pseudofermion chirality. This gives
a very noisy estimator for $\Delta S$, with a large mean and deviation.

Changes ``up'' ($Q=0 \rightarrow Q=1$) and ``down'' ($Q=1 \rightarrow Q=0$)
for the fixed chirality algorithm (data set ${\bf B}$) are illustrated in Fig. \ref{fig:updown}.
These distributions are also quite asymmetric.
A breakdown of Fig. \ref{fig:samediff}a would duplicate this figure.
\FIGURE{
\includegraphics[width=0.75\textwidth,clip]{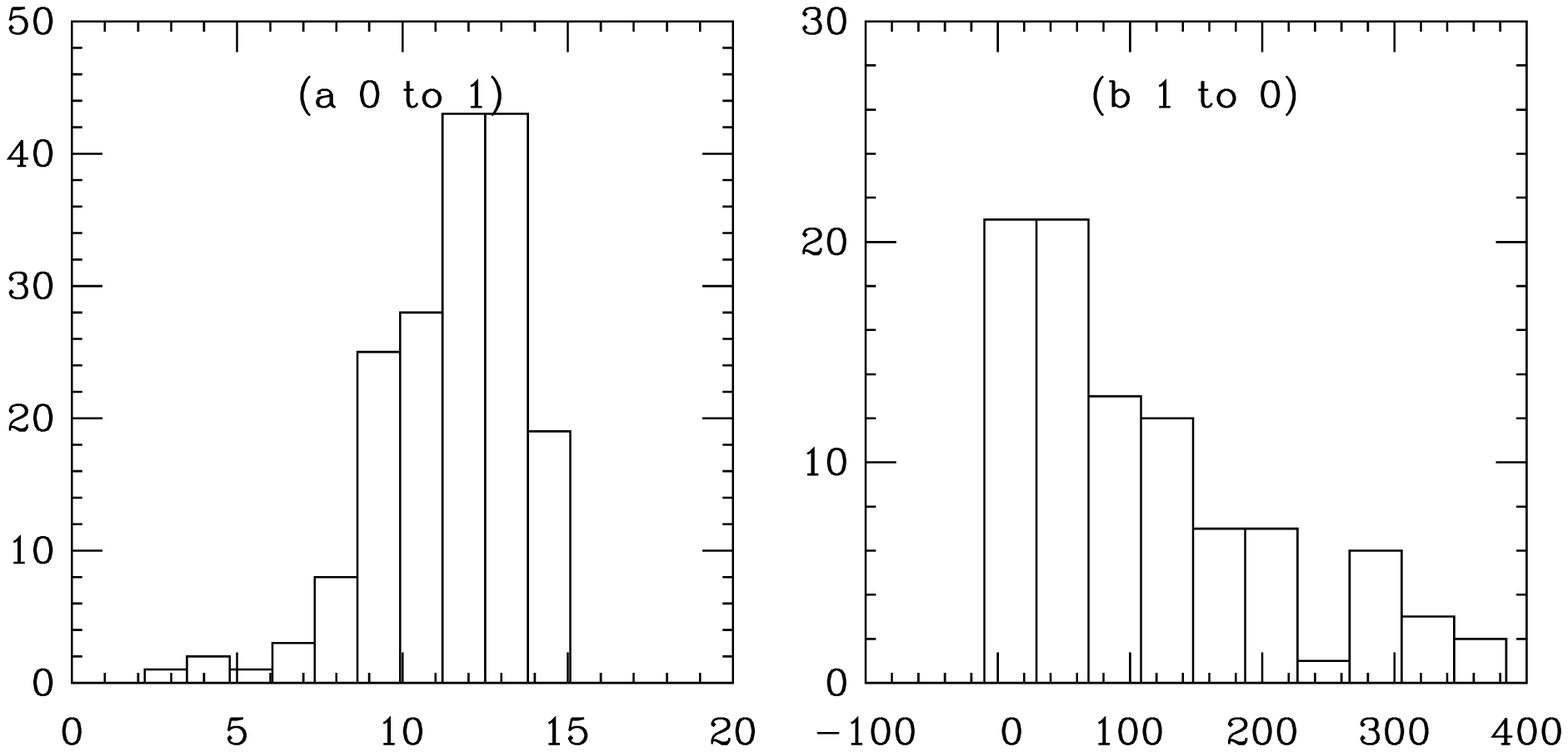}
\caption{Histograms of $\Delta S$ from data set ${\bf B}$, where the proposed topology change is 
from (a) 0 to 1 and (b) 1 to 0.
\label{fig:updown}
}
}

The common feature of these plots is that the distribution of $\Delta S$
is large and wide when the lowest eigenvalue of $H^2$ would shrink 
if the topology changed, and is small and narrow when the eigenvalue would
grow. The lowest eigenvalue in the same-chirality sector shrinks when a zero mode
appears. In the opposite-chirality sector, when the magnitude of the topological
charge increases the smallest eigenvalue also increases, and when the magnitude of $|Q|$ drops,
so does the smallest eigenvalue.

This is most apparent in Fig. \ref{fig:updown}.
The distribution is wide in Fig. \ref{fig:deltas}a 
 because  when we attempt to tunnel out of $Q=0$, a near-zero mode appears in the spectrum.
Fig. \ref{fig:samediff}a contains many low $\Delta S$ values ($Q=0$ to $Q=1)$ and a few
high values ($Q=1$ to $Q=0$) as in Fig.  \ref{fig:updown}. The two
 components in Fig.  \ref{fig:deltas}b are a narrow one for transitions from $Q=0$ to $Q=1$
with $Q=-1$ sources and  a wider one, one for $Q=1$ to $Q=0$ transitions.
 This second component
has a contribution in which the $Q=0$ sector's chirality is flipped.
We don't have enough data from the chiral algorithms
for transitions from $Q=1$ to $Q=2$ to make a histogram,
but there is a strong
hint of a small $\Delta S$ for transitions from 1 to 2
 and a big $\Delta S$ for transitions down.

In all of these distributions, the size of the fluctuations in $\Delta S$ 
for a particular kind of proposed topological change is strongly correlated to the size
of $\Delta S$ itself. This correlation arises
 because $\Delta S$ comes from an average over a set of Gaussian random vectors
and because the change in the pseudofermion action is limited to a single
crossing mode. (We are also assuming that the initial and final pseudofermion
chiralities are identical.)
Taking one pseudofermion and
assuming that we have refreshed immediately before encountering a crossing, the action change per flavor
is
\bee
\Delta S = \xi_\sigma^\dagger( \sqrt{H_\sigma^2}\frac{1}{\tilde H_\sigma^2
}\sqrt{H_\sigma^2} -1) \xi_\sigma
\ee
where  $1/H_\sigma^2$ is given by Eq. \ref{eq:dspf}.
The average is done with respect to a weight factor
which is a pure Gaussian,
\bee
\langle O \rangle = \frac{ \int d\xi^\dagger  d\xi\ O\ \exp(-\xi^\dagger  \xi)}
{\int d\xi^\dagger  d\xi \exp(-\xi^\dagger  \xi)},
\ee
so if $O \sim \xi^\dagger  V \xi$, then $\langle O \rangle = \Tr V$. The trace runs over one 
state, the crossing state, and so $\Tr V$ is the number $V_0$ or
\bee
\langle \Delta S \rangle = -\frac{\delta C L}{1 + \delta C L}.
\ee

Because the measure is just a Gaussian, 
 the squared variance is
 $\sigma^2 = \langle (\Delta S)^2\rangle - \langle \Delta S\rangle^2 = \Tr V^2$.
Again,  only the crossing state contributes to $V$, $\Tr V^2 = V_0^2$,
and so the variance $\sigma$ is equal to the absolute value of $\langle \Delta S\rangle$.

Because a topological boundary can only be crossed when
 $\langle N\cdot H\rangle^2 > 2 \Delta S$,
and because $\langle N\cdot H\rangle^2$ is always on the order of unity
(recall Fig. \ref{fig:ndoth}), only when $\Delta S$ is in its low-value tail
can a crossing occur. However, because the average value of $\Delta S$ is equal to
its fluctuation, this is a constant fraction of the $\Delta S$ sample.
This is how the algorithm preserves detailed balance.

\section{A model calculation}

To illustrate our results, we have constructed a solvable model 
with a discontinuity in its spectrum.
It is a simple system which is confined to a box and 
inside of that box has two regions with different weights. To be specific,
the partition function is
\bee
Z= \int dx
\begin{cases}
\infty & x < -1 \\
\det M_L^2 & -1< x < 0 \\
\det M_R^2& 0 < x < 1 \\
\infty & x > 1
\end{cases}
\ee

For $x<0$, the weight is given by the determinant of
\bee
M_L=\left(\begin{array}{cc} 1+m & 1 \\ 1 & 1+m  \end{array}  \right)
\ee
and for $x>0$ simply by
\bee
M_R=\left(\begin{array}{cc} 1+m & 0 \\ 0 & 1+m  \end{array} \right) .
\ee
Besides having different determinants, the
 matrices do not commute, so their eigenvectors change across the step at $x=0$.
Since $M_L$ has eigenvalues $m$ and $2+m$ it plays the role
of the sector of QCD with the lower eigenvalue of the Dirac operator.

We simulate this theory with the HMC algorithm of Ref. \cite{Fodor:2003bh}, reflecting off
the walls and reflecting/refracting at the step at $x=0$.
We introduce the determinant(s)
by pseudo-fermions
\bee
(\det M_i)^2= \int d\phi d\phi^\dagger \exp(-\phi^\dagger(M_i^\dagger
M_i)^{-1}\phi) .
\ee
We allow for the possibility of  Hasenbusch preconditioning.
For $n$ pseudofermion masses, the $j$th mass is $m_j= m^{(n-j+1)/n}$. 

The ``gauge field''  is the position variable $x$ which we drive with a momentum $p$.
We want to see two things: (i) What is the width of $\Delta S$ in barrier crossings?
(ii) What affects the tunneling rate?
Fig. \ref{fig:ds1} shows $\Delta S$ at the crossing from a simulation  in the model
with $m=0.1$ and a single pseudofermion. The agreement with what we saw in the QCD
simulation -- a wide distribution when the minimum eigenvalue drops, a narrow
distribution when it rises --  is striking.

\FIGURE{
\includegraphics[width=0.8\textwidth,clip]{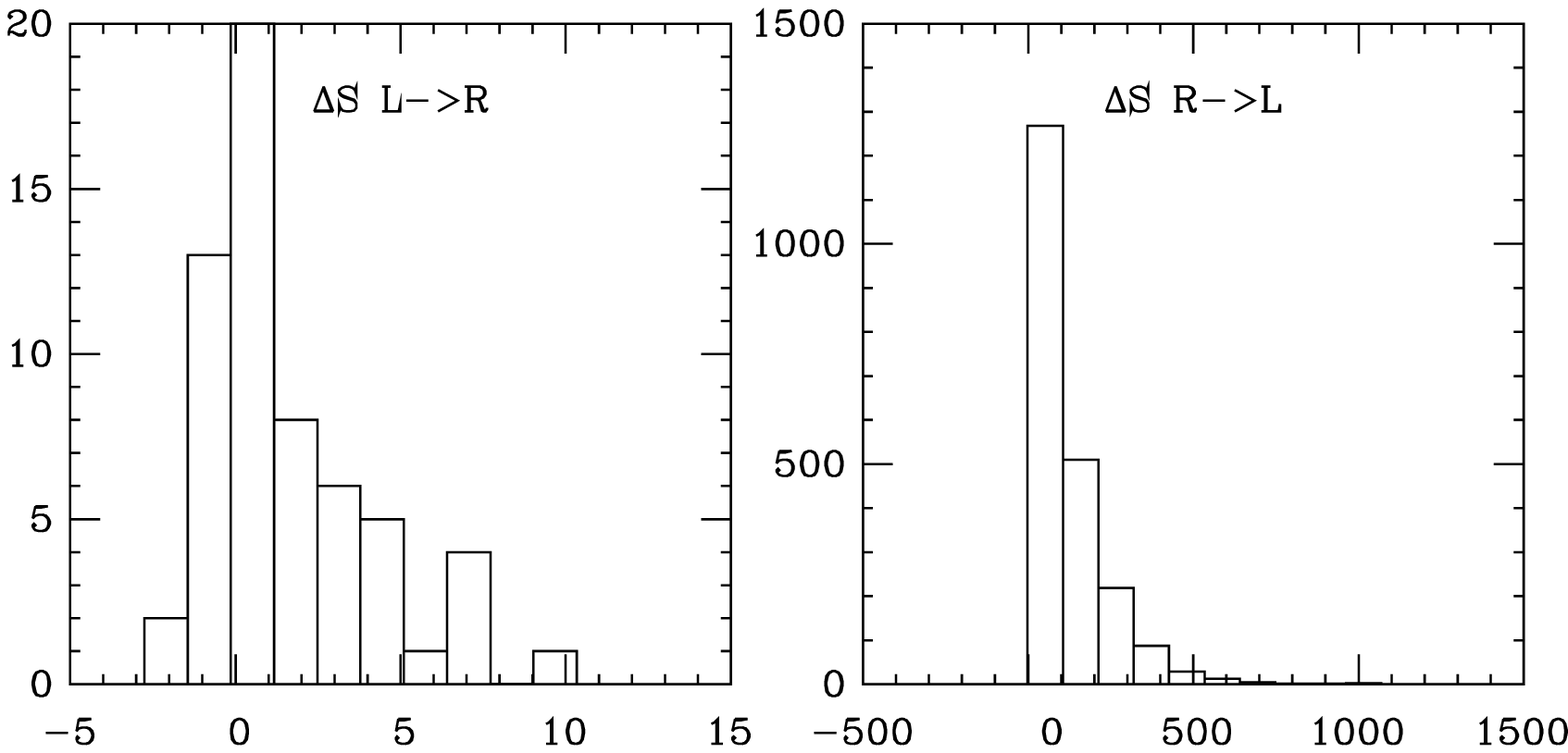}
\caption{$\Delta S$ at the crossing in the model, with $m=0.1$ and a single pseudofermion.
\label{fig:ds1}
}
}

Now for the tunneling rate: we first compute the refraction probability in this
model.
In the exact case, the rate of tunnels from right to left is
\bee
P(R\rightarrow L) =
 N_{LR} \int_0^\infty p dp \exp(-p^2)\theta(p^2 - \log(\det(M_R/M_L)^2) )
\ee
where $N_{LR}$ is the number of left moving particles on the right side
and  the extra factor of $p$ in the integrand counts the flux across the barrier.
The tunneling rate in the other direction is
\bee
P(L\rightarrow R) = N_{RL} \int_0^\infty p dp \exp(-p^2)
\label{eq:exactRL}
\ee
where $N_{RL}$ is the number of right movers on the left side.
The theta-function is absent because the logarithm is always negative,
and the integral is just unity: the crossing rate is 100 per cent. Of course,
$N_{LR}=N_{RR}$ and $N_{RL}=N_{LL}$ from reflections at the ends.
Equating the tunneling rates gives
 $N_{LR} =N_{RL}\det(M_L/M_R)^2$ which is the statement of detailed balance.

Eq.~\ref{eq:exactRL} represents an upper bound on the tunneling rate from right to 
left, and the tunneling rate for HMC will always be less than this value.
Since detailed balance is obeyed, the tunneling rate in the other
 direction is also suppressed.

In the stochastic case and if we are on the left side,
 we choose $\phi = M_L^\dagger \xi$ where $\xi$ is Gaussian.
 Then $\Delta S = \xi^\dagger(W -1) \xi$ where $W=  M_L(M_R^\dagger M_R)^{-1}M_L^\dagger$.
We can rotate the  pseudofermion
integration measure to a basis which diagonalizes $W$
and the tunneling rate is
\bee
P(L\rightarrow R) = N_{RL}  \int_0^\infty p dp \exp(-p^2) \prod_{i=1}^2 d\xi_i^2
\exp(-\sum_i\xi_i^2) 
\theta(p^2- \sum_i \xi_i^2(\lambda_i -1) ).
\ee
If the eigenvalues of $W$ were all less than unity, the momentum integral would
be unconstrained. That does not happen, however: it is easy to show
that the eigenvalues of $W-1$ are $\epsilon_1=(3+2m)/(1+m)^2$ and
$\epsilon_2= -(1+2m)/(1+m)^2$. The tunneling rate is reduced from
$N_{LR}$ to $N_{LR}[1 - \epsilon_1^2/(1+\epsilon_1)(\epsilon_1-\epsilon_2)]$.
To preserve detailed balance, the tunneling rate in the opposite direction must be suppressed
by the same amount.

Next we add $n$ extra Hasenbusch pseudofermions. The matrix $W_n$ 
(for the heaviest pseudofermion) is identical to what we computed above;
 for the lighter pseudofermions, 
\bee
W_j = M_L(m_{j}) M_L^\dagger(m_{j+1})^{-1} M_R^\dagger(m_{j+1)}
 (M_R^\dagger(m_j)M_R(m_j))^{-1}   M_R(m_{j+1})  M_L^\dagger(m_{j+1})^{-1} M_L^\dagger(m_{j}) .
\ee
The theta function becomes $\theta(p^2 - \sum_{ij}|\xi_{ij}|^2(\lambda_{ij}-1))$.
An analytic formula is unilluminating. However, one discovers the
following result: One of the eigenvalues of $W_j$ is always less than unity. The other one
is greater than unity, but as the number  of pseudofermions increases,
this eigenvalue falls to a value which is only slightly greater than unity.
Thus the constraint on the lower end of the
momentum integral relaxes and the tunneling rate rises toward unity,
the deterministic result. 
(For example, for $n=8$ and $m=0.05$
 the lower mass pseudofermions' largest eigenvalue ranges from 1.03 to 1.11. The highest
pseudofermion has a largest eigenvalue of 2.5. For one pseudofermion the one relevant eigenvalue
is equal to 3.8.)
More pseudofermions enhance the tunneling rate.

A graph of this behavior from a series of simulations
 is shown in Fig. \ref{fig:ltor0.05} for $m=0.05$.
Most of the change happens with the first few pseudofermions. We are not sure whether
this result has much practical use in QCD, since our algorithm is costly enough that
many pseudofermions are simply too expensive.
\FIGURE{
\includegraphics[width=0.8\textwidth,clip]{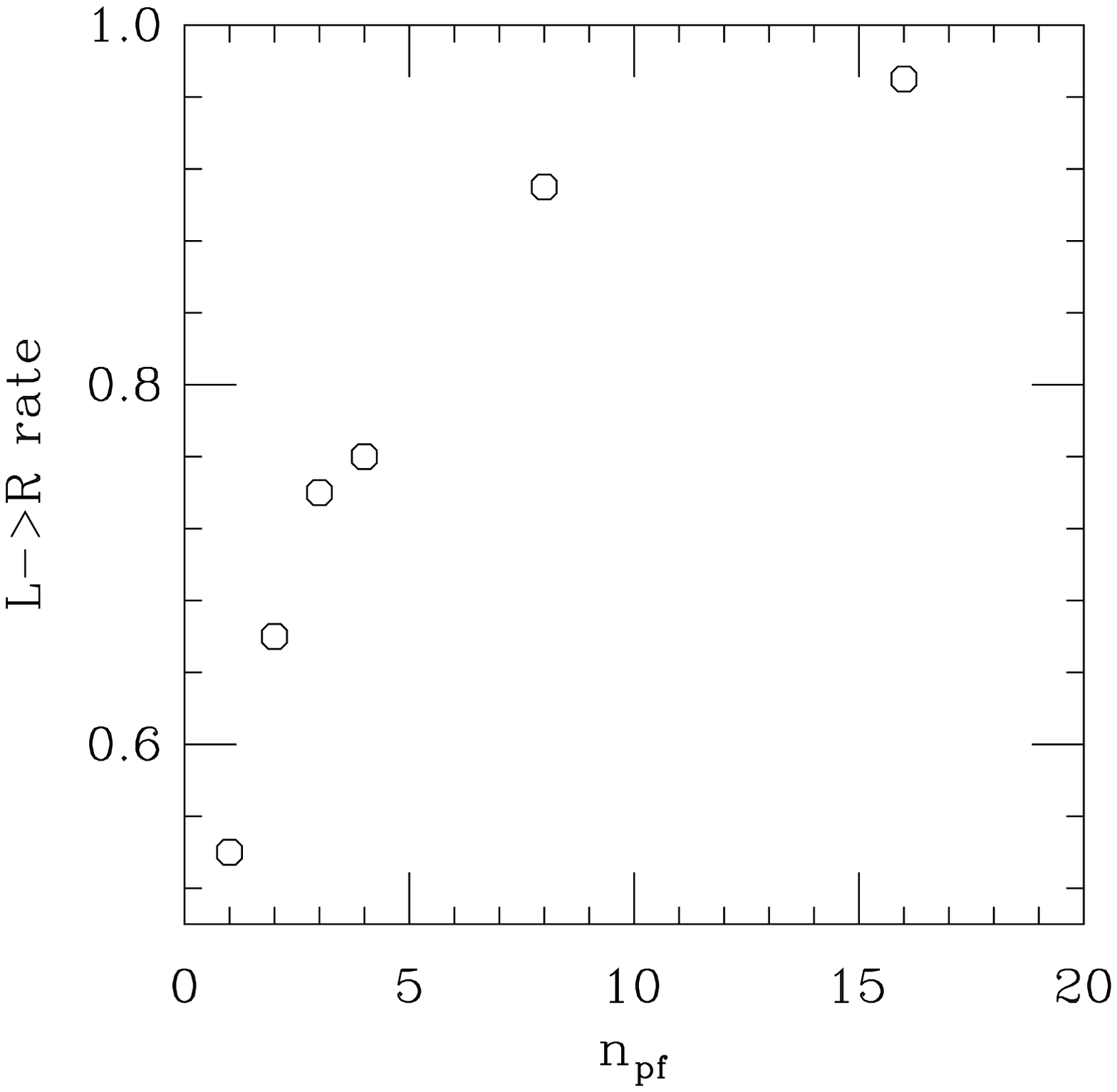}
\caption{Crossing rate from left to right at the discontinuity from a simulation of the model
 for $m=0.05$.
\label{fig:ltor0.05}
}
}

The average (stochastic) $\Delta S$ is related to its variance as above.
The eigenvalues of $W$ also give us the value of $\Delta S$ at the crossing point:
for $n$ pseudofermions the result is
\bee
\Delta S_{L\rightarrow R} = \sum_n \sum_i (\lambda_{ni}-1)
\ee
and
\bee
\Delta S_{R\rightarrow L} = \sum_n \sum_i (\lambda_{ni}^{-1}-1).
\ee
Thus with more pseudofermions, as $\lambda_{ni}$ falls toward 1,
 the means of both distributions shrink.
This is clearly illustrated with a plot of $\Delta S$ for eight pseudofermions, Fig. \ref{fig:ds8}.

\FIGURE{
\includegraphics[width=0.8\textwidth,clip]{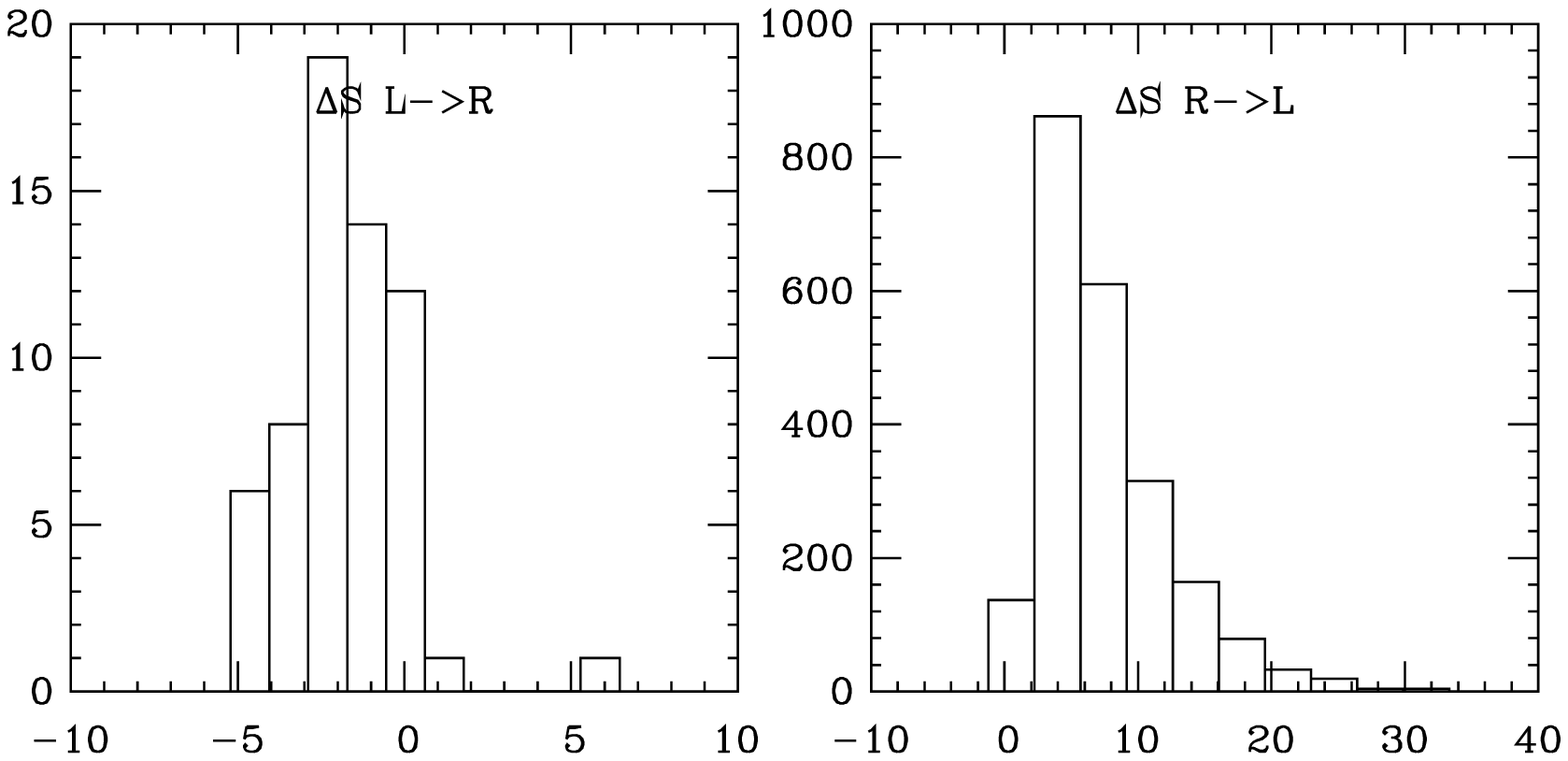}
\caption{$\Delta S$ at the crossing in the model, with $m=0.1$ and 8 pseudofermions.
\label{fig:ds8}
}
}

The conclusion of this model study is that the asymmetric distributions we
 observe do not affect detailed balance. The value of $\Delta S$ is an indirect measure
of the expected tunneling rate, through
 the eigenvalues of the ratio of pseudofermion matrices
at the crossing point. More pseudofermions should enhance the refraction rate.

The model does not directly address the question of whether the chiral algorithm
should have a higher tunneling rate than the nonchiral algorithm.
However, in simulations restricted to the opposite chirality sector
the shift in the spectrum is smaller than in the same chirality sector.
This amounts to a larger value of $m$ in the model. With bigger $m$ the tunneling rate
even with a single pseudofermion is closer to its deterministic value. 
This plus the use of the exact weight for the zero mode suggests
suggests that the chiral algorithm will evolve more efficiently than the usual
nonchiral HMC.

\section{Conclusion}
We have discussed a method to simulate an arbitrary number of flavors of overlap fermions 
in hybrid Monte Carlo. Besides removing the constraint on flavor number from HMC,
the algorithm has some practical features.
 It avoids the problems in the simulation associated with  zero-modes.
 In sectors of non-vanishing topology, this facilitates and significantly stabilizes
the inversion.

As far as we know, this method is only applicable to simulations with the overlap action, since
it needs  the Ginsparg-Wilson relation to relate the spectrum of $D$ to that of $D^\dagger  D$,
that $D^\dagger  D$ commutes with $\gamma_5$, and that changes of topology can be observed
from zero crossings of the kernel action.

\vspace{0.5cm}
\noindent
{\bf Acknowledgments:\\}
We would like to thank Anna Hasenfratz and Roland Hoffmann for discussions.
This work was supported in part by the US Department of Energy.

\end{document}